\documentclass[aps,prc,12pt,nofootinbib, superscriptaddress]{revtex4-1}
\usepackage{selinput}
\usepackage[T1]{fontenc}
\usepackage{graphicx}
\usepackage{epstopdf}
\usepackage{amsmath,amssymb}
\usepackage{verbatim}
\usepackage[dvipsnames]{xcolor}
\usepackage{geometry}
\usepackage{units}
\usepackage{systeme}
\usepackage{amsmath}
\usepackage[toc,page]{appendix}\usepackage{float}
\usepackage{subcaption}
\usepackage{lipsum}
\usepackage{hyperref}
\usepackage{caption}
\newcommand \beq{\begin{equation}}
\newcommand \eeq{\end{equation}}

\begin{document}
\title{Quarkyonic Meson Matter for Finite Isospin Density\footnote{To be published in a special volume of Acta Physica Polonica B dedicated to Andrzej Bialas on the occasion of his 90'th birthday.}}

\author{Larry McLerran}
\affiliation{Institute for Nuclear Theory, University of Washington, Box 351550, Seattle, WA 98195, USA}

\newcommand{\MN}[1]{\textcolor{magenta}{#1}}
\newcommand{\MB}[1]{\textcolor{blue}{#1}}
\newcommand{\TK}[1]{\textcolor{MidnightBlue}{#1}}
\newcommand{\YF}[1]{\textcolor{magenta}{#1}}
\newcommand{\LM}[1]{\textcolor{Red}{#1}}

\newcommand{\kFB}{k_{\rm FB}}
\newcommand{\EF}{E_{\rm F}}
\newcommand{\kFQ}{k_{\rm FQ}}
\newcommand{\kbu}{{k_{\rm bu}}}
\newcommand{\ksh}{{k_{\rm sh}}}
\newcommand{\qbu}{{q_{\rm bu}}}
\newcommand{\qsh}{{q_{\rm sh}}}
\newcommand{\Nc}{N_{\rm c}}
\newcommand{\DB}{\Delta_{\rm B}}
\newcommand{\DQ}{\Delta_{\rm Q}}
\newcommand{\fFD}{f_{\rm FD}}
\newcommand{\fB}{f_{\rm B}}
\newcommand{\betaid}{\beta_{\rm id}}
\newcommand{\muid}{\mu_{\rm id}}
\newcommand{\Tid}{T_{\rm id}}
\newcommand{\Tr}{\mathop{\mathrm{Tr}}}
\renewcommand{\Re}{\mathop{\mathrm{Re}}}

\begin{abstract}
QCD at finite isospin density is considered for a large number of colors $N_c$. A linear sigma model is used to model the meson content of the theory at low density.  At isospin chemical potential $\mu_I << \Lambda_{QCD}$, this matter forms a Bose condensate.  For $\mu_I >> \sqrt{N_c} \Lambda_{QCD}$, unlike QCD remains confined, but the degrees of freedom of the system are mesons and Cooper pairs bound on size scales small compared to the QCD size scale  determined by the superfluid gap.  For most purposes this matter may be analyzed using weak coupling methods.   For $ \Lambda_{QCD} \le \mu_I \le \sqrt{Nc} \Lambda_{QCD}$, we argue that meson matter is quarkyonic, with quarks bound into mesons on a size scale of order $\Lambda_{QCD}$ corresponding to a filled Fermi sea of quarks, with possible Bose condensation at the Fermi surface and/or Cooper pairs with finite width of the surface of order $\Lambda_{QCD}$.

\end{abstract}

\maketitle

\section{Introduction}

QCD at finite isospin density was originally proposed as a theory to describe non-perturbative aspects of QCD  at finite density\cite{Son:2000xc}.  It has been studied on the lattice \cite{Kogut:2002tm}\cite{Kogut:2002se}\cite{Brandt:2017zck}\cite{Brandt:2022hwy}\cite{Abbott:2023coj}\cite{Abbott:2024vhj}, and more recently using analytic methods within the non-linear sigma model\cite{Ayala:2023cnt}.  Lattice computations are straightforward for finite isospin density because there is no sign problem for the fermion determinant\cite{Son:2000xc}. The main features of the phase diagram are well known.  

At very low densities, the system remains in vacuum  until the
chemical potential reaches the pion mass.  The system is vacuum with a non-zero chiral condensate.  Above the pion mass the orientation of the chiral condensate evolves into a condensate along the direction of the chemical potential and the matter becomes a pion condensate.

At very high densities, the Debye mass becomes larger than the confinement scale, if we ignore superfluidity effects, the system would become a Fermi gas of up and anti-down quarks, assuming that the isospin chemical potential is $\vec{\mu}^i = \mu_I \delta^{i3}$.  It is however known that  a superfluid condensate forms\cite{Alford:1997zt}\cite{Schafer:1996wv}.   Such a condensate affects the confining nature of the theory\cite{Son:2000xc}, and the scale where this density is achieved is presumably the density scale at which conventional computations of Cooper pairing  might be reliable.  At finite isospin density, the contribution of this condensate is not exponentially suppressed at large $N_c$. It is also inferred from lattice computations\cite{Abbott:2023coj}\cite{Abbott:2024vhj} that such a condensate might make a significant contribution to the pressure or energy density at very high isospin density.\cite{Fujimoto:2023mvc}.  This condensate generates a gap in the quark spectrum, and this gap implies that the system is in fact confining\cite{Son:2000xc}.  The resulting theory is a theory  of bound mesons and Cooper pairs with the typical distance scale of bound states and Cooper pairs determined by the gap, $\Delta$.  Since the gap, $\Delta >> \Lambda_{QCD}$,  the dynamics is for most purposes controlled by a weak coupling, and computable using weak coupling methods.  Nevertheless, properties of the system measured at long distances exhibit confinement and the excitations of the Fermi surface are confined. 

The typical momentum scales at which these different physics effects takes place is controlled by the isospin chemical potential.  For $\mu_I \le \Lambda_{QCD}$, the dynamics is non-perturbative, and the degrees of freedom are mesons.  On the other hand, the effectively weakly coupled phase which is controlled by the gap associated with superfluidity, requires that the Debye mass $M_{Debye} >> \Lambda_{QCD}$.  This is necessary for the superfluid computation so that weak coupling methods may be used in the analysis\cite{Son_1999}.  This is presumably a density where the gap scale becomes larger than the confinement scale, and the dynamics is controlled by a weak coupling.  At large numbers of colors, $M_{Debye}^2 \sim \alpha_{'t Hooft} \mu_I^2/N_c$,
since Debye screening is generated by a quark loop.  Here $\alpha_{'tHooft} =
g^2 N_c/(4\pi)$ where $g$ is the gauge coupling.  Therefore there is some intermediate density range where $\Lambda_{QCD} << \mu_I << \sqrt{N_c} \Lambda_{QCD}$, where the dynamics would be naively weakly coupled but for which processes sensitive to the infrared, such as Cooper pairing are essentially non-perturbative.  Presumably quarks  and Cooper pairs are bound in mesons of size scale $1/\Lambda_{QCD}$, and the dynamics is quite different from that
at higher densities, where the size scale of mesons and Cooper pairs is of order $1/\Delta$.   

In the analysis below, we will study QCD with finite isospin density and zero baryon number density in the limit where the pion mass is vanishingly small, and the number of colors is large.  
We will see that the transition between the low density confined phase and the high density phase controlled by weak coupling is  similar to what occurs with quarkyonic matter at finite baryon number density.   We will call this intermediate region region  $\Lambda_{QCD} << \mu_I << \sqrt{N_c} \Lambda_{QCD}$ where matter may be for some purposes treated as mesons and some purposes as quarks as quarkyonic meson matter.  We can think of this matter in analogy to quarkyonic baryon matter as a filled sea of quark with a non-perturbative surface of mesons.  We will also see that this quakryoninc meson matter has a dual description both in terms of either mesons or quarks\cite{Fujimoto:2023mzy}.

Quarkyonic matter occurs at finite baryon number density and large $N_c$ for baryon chemical potential $\mu_Q \le \sqrt{N_c} \Lambda_{QCD}$, where $\mu_Q = \mu_B/N_c$ is the quark chemical potential corresponding to the baryon chemical potential $\mu_B$\cite{McLerran:2007qj}.  This is because the effects of Debye screening are suppressed, $M_D^2 \sim \alpha_{'t Hooft} \mu_Q^2/N_c$.  The Debye mass is less than the confinement scale $\Lambda_{QCD}$ and so the system remains confining until a large chemical potential is achieved.  The Debye screening mass becomes larger than the confinement scale when $\mu_Q^2 \ge N_c\Lambda_{QCD}^2$, and there is therefore an intermediate range of densities $\Lambda_{QCD}^2 << \mu_Q^2 << N_c \Lambda_{QCD}^2$ where matter has weak coupling for many purposes, and for these purposes may be thought of as perturbative, but is nevertheless confined on a scale of ordere\ $\Lambda_{QCD}$.
In this intermediate regime, confinement effects are important when physics is probed at scales large compared to the confinement scale.  It was argued that matter at these intermediate density scales might be effectively thought of as a Fermi surface composed of nucleons surrounding a filled Fermi sea of quarks\cite{McLerran:2007qj}.  An explicit formulation which has dual description in terms of baryons and nucleons was constructed in Ref. \cite{Fujimoto:2023mzy}.  In terms of nucleons, the baryon number distribution corresponds to a shell of nucleons of full occupation number surrounding a sea of under occupied baryons, with occupation number $\sim~ 1/N_c^3$.  This corresponds to a fully occupied sea of quarks with an exponentially falling tail of quarks at the Fermi surface.

A similar phenomenon might happen in QCD at zero baryon number density and finite isospin chemical potential.    This means that matter makes a transition to a confined but largely weakly coupled phases when $\mu_I^2 \sim N_c \Lambda^2_{QCD}$.  On the other hand at the much lower scale, when $\Lambda_{QCD}^2 << \mu_I^2  <<N_c \Lambda^2_{QCD}$, we expect that for typical distance scales, the effect of interactions should be approximately perturbative.  Nevertheless interactions at the Fermi surface can become strong and not analyzable with weak coupling methods where interactions on the scale of order $\Lambda_{QCD}$ may become important.  Therefore at finite isospin density, there is an intermediate phase of matter in direct correspondence to quarkyonic matter that appears at finite baryon density.  A major difference is that here the relevant hadronic degrees of freedom are mesons not baryons.

In what follows, we will argue that in this intermediate density region, there is a dual description between mesons and quarks.  Qualitatively, matter may be thought of as a  filled Fermi sea of quarks surrounded by a shell which is a Bose condensate of pions.  This Bose condensate has two potential contributions.  The first is like what happens for quarkyonic baryonic matter. A highly occupied pion condensate is formed at a momentum of twice the Fermi momentum for the $u$ or $\overline d$ quarks.  This condensate
may be thought of as the analog of charge density waves.  This is a delta function contribution in the momentum of the pions and the occupation number of this condensate is of order $N_c$.  There is another potential contribution which may arise from Cooper pairs.  Cooper pairs have a different wavefunction than a pion. The $u$ and $\overline d$ quarks have momentum distributions centered around
$\mid k_u \mid \sim \mid k_{\overline d} \mid  \sim k^F$, where $k_F$ is the typical momentum of a Fermi surface for $u$ or $\overline d$ quarks.  The Cooper pairs have zero total momentum and form  a spatially homogeneous condensate

We will see that the  reason for the formation of these two condensates is to fill up the momentum distribution of quarks near the Fermi surface which are not quite fully occupied in the absence of either of the condensate contributions.  This can either be filled by the finite momentum space width quark distributions corresponding to mesons at the Fermi surface with momentum $2k^F$, or with the spread in momentum around $k^F$ of the quarks inside a zero momentum Cooper pairs.  In either case the resulting Fermi distribution for quarks will have a tail corresponding to the momentum space widths of either of these two contributions.

\section{Filling up with KFC}

The formation of quarkyonic matter at finite baryon density is signaled by a change in the baryon number distribution. If one computes the distribution of quarks arising from a Fermi distribution of baryons, one finds  that for a conventional Fermi distribution, the density of quarks becomes over-occupied. This may be seen from the Kojo Filling Criteria (KFC)\cite{Kojo:2021ugu}, which follows from the duality relation that expresses baryon number phase space distribution of quarks $\rho_Q$   in terms of the phase space of nucleons determined by the distribution of baryon number $\rho_N$,
\begin{equation}
          \rho_Q (\vec{k}) = \int {{d^3p} \over {(2\pi)^3}} K(\vec{k}-\vec{p}/N_c)~\rho_N(\vec{p}) 
          \label{eqn:duality}
\end{equation}
where $K(k)$ is the normalized probability distribution to see a quark of momentum $k$ in a nucleon at rest,
\begin{equation}
    \int {{d^3k} \over {(2\pi)^3}} K(\vec{k}) = 1
\end{equation}
The phase space distribution of nucleons for a free Fermi gas is $\rho_N(\vec{p}) = 1$ for $p \le p^F$.
The criterion that $\rho_Q(\vec{k}) < 1$ leads to the KFC
\begin{equation}
    \int {{d^3p} \over {(2\pi)^3}} K(\vec{p}/N_c)~\rho_N(\vec{p}) \le 1
\end{equation}
This relationship implies that there is a maximum possible density for a Fermi distribution of nucleons where the KFC is satisfied.  Above this critical density, the nucleon distribution must be different.  This region of high density with a modified nucleon distribution is quarkyonic.

At the onset of quarkyonic region,  there is no Fermi sea for the quarks, and the distribution of quarks arising from the baryons should be maximal at $k = 0$ with a value less than 1.  The onset of quarkyonic matter is when this density of quarks saturates at $k = 0$
If the maximum momentum of the distribution of quarks is much larger than the width of the probability distribution of nucleons, and if the quark baryon number distribution is saturated, this relationship requires that
\begin{equation}
    1 = \int {{d^3p} \over {(2\pi)^3}} K(\vec{p}/N_c) \rho_N(\vec{p}) = N_c^3 \rho_N^0
\end{equation}
assuming that the nucleon baryon number distribution is constant for 
\begin{equation}
    \rho_N ( \vec{p}) = \rho_N^0, ~~~~ p << p_{surface}
\end{equation}
Here, $p_{surface}$ is the limiting momentum of the nucleon momentum distribution.  This requires that for maximum quark occupancy, the baryons are under-occupied, $\rho_N^0 = 1/N_c^3$.  The energy is further minimized if at the surface momentum one adds a surface shell of nucleons. This allows us to patch on a tail of the quark distribution that
falls from one to zero on a scale associated with the distribution of quark momenta in a nucleon.

This is necessary if quarks are confined, since a sharp distribution of nucleons will always lead to a distribution for quarks with a falling tail, due to the intrinsic motion of quarks inside a nucleon.  We see however that this falling tail of the quark phase space distribution corresponds to a nucleon sitting on the Fermi surface.  The reason why this is consistent with the KFC, is that near the Fermi surface, the part of the nucleon distribution corresponding to under-occupation leads to a falling distribution of quarks.  This can be filled in for momenta less than the surface momenta by a localized distribution of nucleons with maximum occupation number.  For quark momenta larger than that of the surface momenta, the quark distribution falls exponentially, corresponding to the nucleons of ordinary occupation number that sit near the Fermi surface.

\begin{figure}
  \centering
  \vspace{-0.5cm}
  \includegraphics[width=0.95\columnwidth]{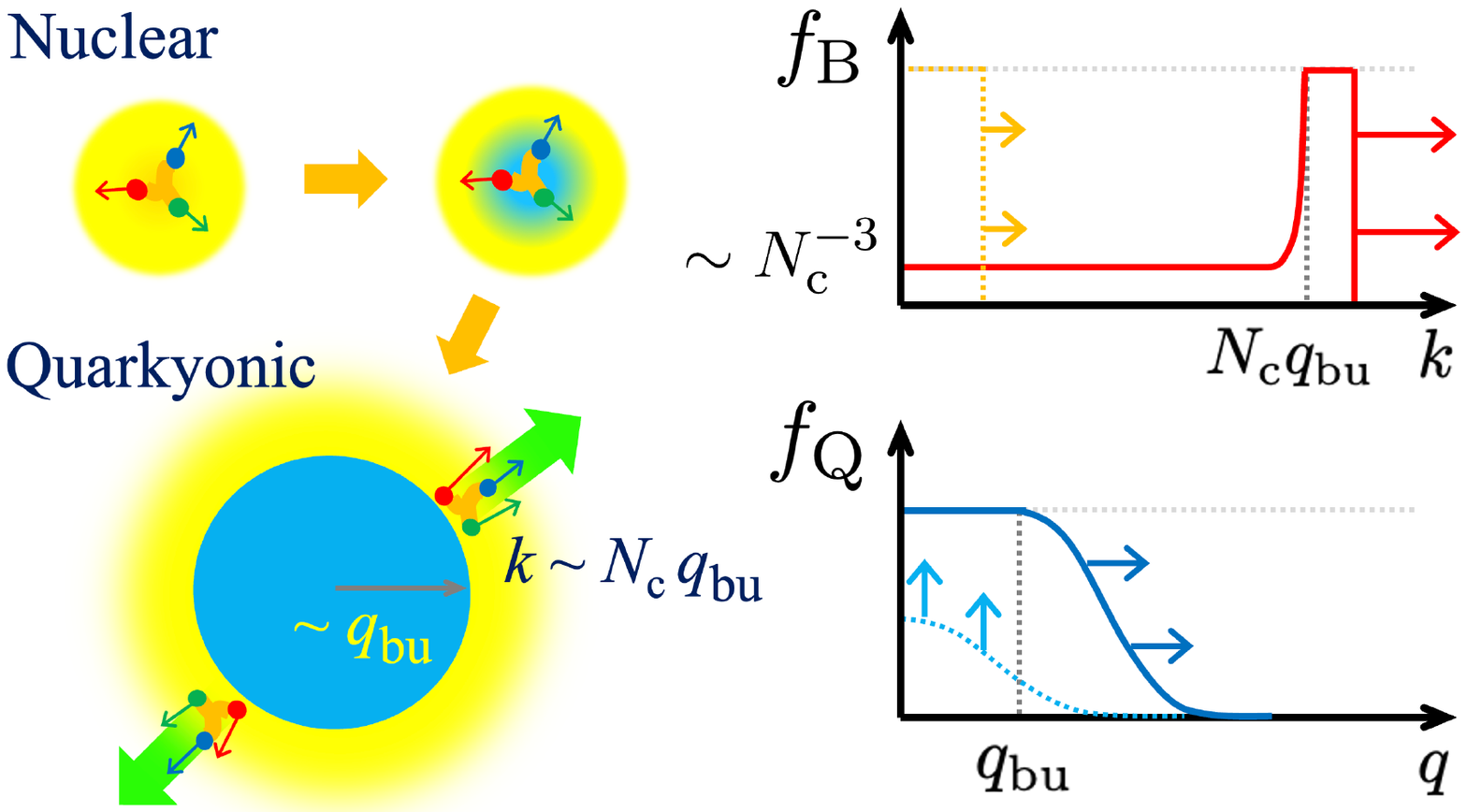}
    \vspace{-0.5cm}
  \caption{Evolution of $f_B$ and $f_Q$ from the nuclear (dotted) to quarkyonic regime (solid).
  The saturation of quark states drives baryons into the relativistic regime.
  }
    \vspace{-0.5cm}
  \label{fig:shell}
\end{figure}

Quarkyonic matter and the distribution of quarks and nucleons is illustrated in Fig. \ref{fig:shell}.  Note that the typical scale of momentum in the quark baryon number distribution is $1/N_c$ relative to the typical scales of the nucleon baryon number distribution.  This is because $N_c$ quarks inside a nucleon share the momentum of the nucleon.  This is a consequence of Eqn. \ref{eqn:duality}

The system at high isospin density can be analyzed by similar methods.  The KFC relationship for isospin is slightly different.  We will assume that the finite isospin density is generated by an excess of  mesons composed  of a $u$ quark and an anti $\overline d$.  We will consider simultaneously the small mass pion limit, and a large number of colors.  At the lower densities, these mesons are pions and should be described by a $\sigma$ model.

The KFC relation for finite isospin is given by recalling that a pion is composed of a $u$ quark and anti 
$\overline d$.  
Therefore the phase space density of quarks, 
$\rho_Q^I = \rho_u^I = \rho_{\overline{d}}^I $ can be expressed in terms of the meson phase space density of isospin
\begin{equation}
    \rho_Q^I(\vec{k}) = {1 \over {2N_c}} \int {{d^3p} \over {(2\pi)^3}} K_M(\vec{k}-\vec{p}/2) \rho_M^I(\vec{p})
\end{equation}
where $K_M$ is the normalized distribution to find a quark inside of a meson.  This is presumably dominated by pions.
The factor of $1/2N_c$ is arises from the spin-color degeneracy of quarks.  The factor of $p/2$ in the integration on the right hand side of the above equation is because each quark carries $1/2$ of the mesons momentum.

We will soon see that in the ordinary meson phase, the isospin density is contained in a Bose Einstein condensate, 
\begin{equation}
\rho_M^I =  (2\pi)^3 \delta^{(3)} (\vec{p}) \rho_C 
\end{equation}
so that the KFC criteria is saturated when 
\begin{equation}
    \rho_C^{saturation} = 2 N_c/K_M(0) = 2N_c {{\Lambda^3} \over {(2\pi)^3}}
\end{equation}
The subscript $C$ denotes Cooper pairs.
$K(0)$  is of order $1/\Lambda^3$
where $\Lambda$ is of the typical momentum space spread of the quark distribution.  I define $\Lambda$ by
\begin{equation}
    K(0) = {{(2\pi)^3 } \over {\Lambda^3}}
\end{equation}

In the sections below we argue how this saturation limit is approached and what are the general features of  the quarkyonic solution at densities higher than this saturation density.

\section{Low Isospin Density}

Consider QCD at finite isospin density for very small mass quarks at large numbers of colors and 2 flavors.  We first consider this theory in the context of the linear sigma model for massless pions. The density at which other mesons appear corresponds to an isospin chemical potential of the order the $\rho$ meson mass, and we will assume that the dynamics that we consider may be largely captured before the $\rho$ meson degrees of freedom become important.

A massive sigma field is condensed in vacuum corresponding to chiral symmetry breaking. The action for the theory is 
\begin{equation}
    S = \int ~d^4x ~\left\{  ~{1 \over 2}~\partial^\nu \Phi^\dagger \partial_\nu \Phi -{{M^2} \over 2}\Phi^\dagger \Phi+ {\lambda \over 4 N_c} (\Phi^\dagger \Phi)^2\right\}
\end{equation}
where $\phi = (\sigma, \vec{\pi})$ is a four component field with an O(4) symmetry.  The meson interaction strength goes as $1/N_c$ in the large $N_c$ limit and $M^2 \sim \Lambda^2_{QCD}$, and $\lambda \sim 1$.
This expression is in Euclidean space.

The chiral symmetry is broken in vacuum by the expectation value of the scalar $\sigma$ field with expectation value,
\begin{equation}
<\sigma > = \left({N_c \over \lambda}  \right)^{1/2 } M
\end{equation}
where $M$ is the mass that appears in the sigma model Lagrangian.
This expecations value  breaks the O(4) rotational invariance, which would allow the system to have a direction anywhere within the $(\sigma, \vec{\pi})$ space.

At finite density, we introduce an isospin chemical potential coupling to the pion field kinetic energy
\begin{equation}
 S_{\pi KE} = {1 \over 2}\int ~d^4x ~\left\{  ~(\partial^\nu +\delta ^{\nu,0}\mu_I T^3) \pi (\partial_\nu -\delta_{\nu, 0}T^3 \mu_I)\pi \right\}
\end{equation}
where $T^3$ is the isospin charge vector component.  We will be concerned with the component independent of spatial gradiants, where for the adjoint representation
\begin{equation}
    T^3_{bc}T^3_{cd} = \delta_\perp^{bd}
\end{equation}
Here $b,d$ are only along the $1,2$ components.  The effective potential is
\begin{equation}
     V = -{M^2 \over 2}(\sigma^2 + \vec{\pi}^2) -{\mu^2_I \over 2} \pi_T^2 + {\lambda \over {4N_c}}(\sigma^2 + \vec{\pi}^2)^2 
\end{equation}

This has an extremum for 
\begin{equation}
    \pi_T^2 ={N_c \over \lambda} (M^2 + \mu_I^2)
\end{equation}
Note that for the case we consider, massless pions, the vacuum is mis-aligned with respect to the minimum at finite $\mu_I$.  This is of course corrected when one allows for finite pion mass, and the alignment changes in the region $\mu_I \sim m_{pion}$.  We are not going to be concerned with this threshold behaviour and will here start from this realigned configuration.

This theory has a conserved current in Minkowski space
\begin{equation}
j^\mu = -{i \over 2} ~\pi_T (\overleftarrow{\partial}^\mu- \overrightarrow{\partial}^\mu) T^3_I\pi_T  +\mu_I \delta^{\mu 0}\pi^2_T 
\end{equation}
or
\begin{equation}
 j^0 = -{i \over 2} ~\pi_T (\overleftarrow{\partial}^0- \overrightarrow{\partial}^0) T_I^3\pi_T  +\mu_I \pi^2_T
\end{equation}
In these equations, the subscript $T$ denotes $1,2$ components of the pion field
\begin{equation}
 V = -{(M^2+\mu^2)\over 2} \pi^2_T - {M^2 \over 2}  \pi^2_3+ {\lambda \over {4N_c}} \pi^4
\end{equation}
and $\pi^4 = (\pi_T^2 + \pi_3^2)^2$.
This follows from $n = dP/d\mu = -dS/d\mu$
This expression for the charge has positive value for a constant expectation value of $\phi$, or a Minkowski space field that oscillates as $e^{-i\omega t}$.  

The minimum of the potential is in the isospin direction, which is misaligned from the value when a small pion mass is present, and when the chemical potential is of the order of the pion mass.  We will consider chemical potential much larger than the pion mass, so that we can ignore to leading order the effect of the pion mass.  
The value of the potential energy at this minimum is
\begin{equation}
 V_0 = -{N_c \over { 4\lambda}} (\mu_I^2+M^2)^2
\end{equation}

Note that this theory has 1 Goldstone mode corresponding to rotations in the $\pi_T$ orientation.  Also, it is interseting to note that the small fluctuations of this action are characterized by a coupling of strength of order $1/N_c$.  It is difficult to see how by computing quantum fluctuations to this scalar theory, we would ever escape the condensate.  Such corrections would be of order $1/N_c$ relative to the condensate term.  Presumably, the resolution to this is that as the chemical potential becomes of the order $\mu_I \sim \Lambda_{QCD}$, then other modes appear, and when the number of these modes is of order $N_c$, then perturbation theory breaks down.  However, before these modes play a dominant role, we will argue that the matter becomes quarkyonic.  We see that from the condensate modes,
\begin{equation}
  \rho_M^I = {N_c \over \lambda}\mu_I(\mu_I^2 + M^2) (2\pi)^3 \delta^{(3)} (\vec{p})
  \label{eqn:phasecondensate}
\end{equation}
and that the KFC is satisfied when $\mu_I \sim \Lambda$.  Therefore Quarkyonic matter forms when $\mu_I \sim \Lambda \sim \Lambda_{QCD}$.  To understand how the momentum space delta function corresponds to the charge density we compute in coordinate space, recall that the Fourier transform of the constant field for the the charge density is a delta function, so for the constant field, the phase space density must  be a delta function in momentum space.  The integral over momentum of the phase space density is the total number of pions, which is equal to the charge in pions, so that the total charge in the condensate is equal to the total number of pion, hence Equation \ref{eqn:phasecondensate}

\section{Expected Features of the Quark and Meson Distributions in the Quarkyonic Region}

For $\Lambda_{QCD}^2 << \mu_I^2 << N_c \Lambda_{QCD}^2$, I speculate on general features of the phase space distribution of mesons and of baryons.  I will assume that the distribution of mesons is in analogy to that for baryons in quarkyonic baryon matter, ie a flat distribution of mesons corresponding to a filled shell of quarks, and a surface condensate contributions:
\begin{equation}
  \rho_M^I(\vec{p}) = \theta (p_{surface} -p)\rho_M^{bulk} +\delta(p_{surface}
    -p)\rho_M^{surface}
\end{equation}
I will first consider the case ignoring a contribution for Cooper airs,  and at the end of this analysis,  will then consider the contribution of such pairs.

For $0 << p << p_{surface}$, requiring that the quark phase space density is 1, gives,
\begin{equation}
    \rho_M^{bulk} = {N_ c \over 4}
\end{equation}
The surface contribution from the condensate is
\begin{equation}
  \rho_Q^I = {1 \over {2N_c}} \int {d\Omega_p} K_M(\vec{k}-\vec{p}_{surface}/2) \rho_M^{surface}
\end{equation}
The contribution from the surface term is localized near the surface.  The reason why the surface term is there is because the bulk contribution will drop below the maximum value of one before reaching the the surface of the quark Fermi distribution.  This allows a surface delta function to fill in the lost part of the quark distribution at minimal cost of energy for the mesons states.

For an exponentially falling probability distribution in 1+1 dimension, it is easy to verify that this distribution is completely filled in for $k < p^{surface}/2$ and for $k > -p^{surface}/2$. The distribution of quarks falls exponentially according to  the probability distribution $K_M(k)$.  For more general probability distributions and in particular in three spatial dimensions, this cancellation will not be exact, but we believe the tendency to fill in the quark distribution.

It is useful to understand this in more detail using a $1+1$ dimensional model for an exponential probability kernal.  In $1+1$ dimensions, there is only one component of spatial momentum, so we dispense with the vector notation.
Here, the normalized $1+1$ dimensional probability kernal is 
\begin{equation}
    K_{1+1}(k) = {\pi \over \Lambda_0} e^{-\mid k \mid/\Lambda_0}
\end{equation}
For the non-quarkyonic condensate mode.
\begin{equation}
    1 = {1 \over {2N_c} } \rho^{sat} {\pi \over \Lambda_0} 
\end{equation}
or
\begin{equation}
    \rho^{sat} = {{2N_c \Lambda_0} \over \pi }
\end{equation}
Now let us assume a quarkyonic phase space distribution for mesons of the form
\begin{equation}
    \rho_M^I = N_c \theta(p_{surface} - \mid p \mid) +(2\pi) \delta(p_{surface} - \mid p \mid)\rho_{surface}
\end{equation}
The contribution for the surface to the quark density is
\begin{equation}
    \rho_Q^{surface}(k) = \rho_{surface} {\pi \over \Lambda_0}(e^{-\mid k - p_{surface}/2\mid} + e^{-\mid k + p_{surface}/2 \mid})
\end{equation}
The contribution from the bulk term in the region where $-p_{surface}/2 < k < p_{surface}/2$ is
\begin{eqnarray}
    \rho_Q^{bulk} & = & {1 \over {4 \Lambda}} \int_{-p_{surface}}^{p_{surface}} dp ~ e^{-\mid k - p/2 \mid/\Lambda_0}  \nonumber \\
     & = & {1 \over {4\Lambda_0}} (   \int_{-p_{surface}}^{2k} dp ~e^{(-k +p/2)/\Lambda_0} + \int_{2k}^{p_{surface}} dp~e^{(-p/2 +k)/\Lambda_0}) \nonumber \\   
     & = & 1 -{1 \over 2}(e^{-(k+p_{surface}/2)/\Lambda_0} +e^{-{((p_{surface}/2-k)}/\Lambda_0})
\end{eqnarray}
We see that if we choose
\begin{equation}
    \rho_{surface} = {1 \over 2}\rho_{sat}
\end{equation}
then for $\mid k \mid < p_{surface}/2$.  the distribution for the quarks will be precisely 1.  Outside this region the quark distribution fall exponentially.  So we see that the surface delta function of mesons cancels the depletion of the quark distribution near the surface momenta arising from the bulk contribution of mesons, at a cost of only adding a mesons at the surface.  This is how the total energy is minimized.

The modification due to Cooper pairs is to add to the meson source term a contribution
\begin{equation}
    \rho_M^I(\vec{p}) = \theta (p_{surface} -p)\rho_M^{bulk} +\delta(p_{surface}
    -p)\rho_M^{surface} +\delta(\vec{p}) \rho_C
\end{equation}
where $\rho_C$ indicate that this contribution comes from Cooper pairs.  The Cooper pair contribution corresponds to zero total momentum, and will modify
\begin{equation}
    \rho_Q^I = {1 \over {2N_c}} \int {d\Omega_p} K_M(\vec{k}-\vec{p}_{surface}/2) ~\rho_M^{surface} + {1 \over {2N_c}} \int {d\Omega_p} K_C(\vec{k}-\vec{p}_{surface}/2)~ \rho_C
\end{equation}
Here $K_C$ is the probability to find a quark inside a Cooper pair, and will have in general a different shape from the corresponding distribution for a meson.  The computation of the relative magnitudes of these various contributions is beyond the scope of this work but they might be determined by Monte-Carlo lattice simulation.

Of course a direct validation of the phase space distributions for mesons and for quarks in this quarkyonic region from a direct lattice computation would be most compelling.  Meson operators are color singlet, and their correlation functions may be constructed, and the corresponding finite density meson distribution extracted.  But of course  great care must be taken in constructing such correlation functions for underlying quark fields.  Quark distributions must be either constructed from operators in a fixed gauge or using link operators to construct gauge invariant distribution functions.
Although these finite density distribution functions form the backbone of perturbative computation their gauge invariant construction and computation on the lattice is somewhat problematic.

The reason why the high density of meson fields form in the bulk is somewhat amusing.  The KFC restricts the growth of the zero momentum condensate at some density.  This means that the  repulsive $\pi_T^4$ in the linear sigma model action no longer compensates the negative contribution from the kinetic energies,
\begin{equation}
    S_{KE} \sim \int d^4x~ {1 \over 2} \pi_T (-\nabla^2 - M^2 - \mu_I^2) \pi_T 
\end{equation}
and modes with sufficiently small momentum will run away to their maximal values consistent with the phase space density of quarks being less than or equal to one.   Of course such condensed modes carry momentum, and once set at one time will evolve in Minkowski space.  But this time evolution is precisely what is need to give a contribution to the isospin charge density.  Also an oscillation corresponding to evolution along the Goldstone mode of $(\pi_1, \pi_2) $ gives a contribution to the third component of the isospin density $\rho_I^3$.

The behavior of the sound velocity as a function of density was a useful indicator of the onset of quarkyonic matter for baryonic matter\cite{McLerran:2018hbz}.  At low densities the sound velocity is small because the system is made of non-relativistic nucleons.  When quarkyonic matter turns on, the system rapidly becomes relativistic, with sound velocity squared approaching $1/3$.  In the process of this onset the sound velocity might exceed 1/3.  For quarkyonic mesonic matter for very small pion mass, the low density sound velocity squared should approach 1.  In the quarkyonic phase it should be close to 1/3.  It may go below 1/3 during the approach if there is a significant contribution from a hadron resonance gas, but if the quarkyonic matter is approached without the significant onset of a hadron resonance gas, we might expect that the sound velocity squared would approach 1/3 from above.

In fact if meson states other than the pion become important, then the computation of the saturated charge density becomes complicated, and spin orientation becomes important.  Presumably this is a story in itself which would need to be unravelled.

A signal for the transition from the quarkyonic mesons phase to the high density weakly coupled phase is given by the fall off of measured Fermi distributions of quarks.  If the matter is quarkyonic, then the fall off at the Fermi surface should be controlled by the width of the meson wavefunction,
$\sim \Lambda_{QCD}$.  If it is controlled by excitation with a width of order the gap, then it is $\sim \Delta$.

\section{Summary and Conclusions}

I have argued that matter at finite isospin density and large $N_C$ may be quarkyonic,  Such matter composed of bosons can be studied using lattice Monte Carlo methods, and the quarkyonic hypothesis might be tested.

\section*{Acknowledgments}
L. M. acknowledges crucial input from Yoshimasa Hidaka, Yuki Fujimoto and Toru Kojo concerning the importance of Cooper pairs in quarkyonic matter, in particular for their importance in filling the quark Fermi sea near the Fermi surface.  Yoshimasa Hidaka's forceful insistence on understanding the confining nature of the high density weak coupled phase, and the potential impact of Cooper pairs in the quarkyonic phase was particularly important. L. M. also thanks Michal Praszalowicz for his suggestions and careful reading of the manuscript. L. M. thanks partial support from the Institute for Nuclear Theory, which is funded in part by the INT’s U.S. Department of Energy grant No. DE-FG02-00ER4113.

\bibliography{references.bib}

\end{document}